\def\r#1#2#3#4#5{\hangindent 10 mm %
\noindent {\sc #1}, 19{#2}, {\it #3}, {\bf #4},
{#5}.}

\def\newpar#1#2{\vskip5mm \begin{center}{\sc \S#1. #2}
\end{center} \nopagebreak}
\def\gesim{\lower 2pt \hbox{$\buildrel > \over \sim$}}
\def\lesim{\lower 2pt \hbox{$\buildrel < \over \sim$}}
\def\ea{{\it et al \/}}
\documentclass[12pt]{article}
\begin{document}
\begin{center}
{\bf Thermal expansion and the equation of state of Ir and Rh} \vskip 5mm

{\sc By M. I. Katsnelson$^\dag$, M. Sigalas$^\ddag$, A. V. Trefilov and
K.~Yu.~Khromov} \vskip 5mm

{\footnotesize Russian Research Centre ``Kurchatov Institute'', Moscow 123182,
Russia \\
$^\dag$Institute of Metal Physics, Ekaterinburg 620219 ,Russia \\
$\ddag$Ames Laboratory, Iowa State University, Ames, IA, 50011, USA}
\vskip5mm

{\sc Abstract}
\end{center}

{\footnotesize The simplest anharmonic characteristics of Ir and Rh are
discussed in the framework of a previously developed simple pseudopotential
model which describes the elastic moduli, phonon spectra and the
lattice heat capacity in the harmonic approximation of these
metals succesfully. The microscopic Gruneisen parameters, the dependences of
the elastic moduli on pressure, the coefficient of thermal expansion and
the equations of state at the finite temperatures have been calculated.
The {\it ab initio \/}
calculations of the energy-band structure and the equation of state for
Ir at $T=0$ have been done to test the model for adequacy at high
pressures. The values of different contributions (zero-point oscillations,
quasiharmonic, etc.) in the considered thermodynamic characteristics of
Ir and Rh are discussed.}

\newpar {1} {Introduction}

As it is stated in the literature (Gornostyrev \ea 1994, Ivanov \ea 1994,
Katsnelson \ea 1996, hereinafter referred to as I), Ir and its
analogue Rh are set off from other FCC metals because of some distinctive
features of their physical properties: unusual deformation-induced failure,
peculiar defect structure, specific temperature dependence of the effective
Debye temperature, etc. Moreover, the situation is not trivial because in the
case of Ir, unlike the most, if not all of the other FCC metals a rather
broad range of properties (elastic moduli, phonon spectra and lattice heat
capacity) can be described in terms of a simple pseudopotential model
(see I and references therein). Even in $sp$-metals the situation is less
favorable: to describe the elastic moduli of Ca allowance must be made for
the singular  contributions to energy which result from the proximity of
the Fermi surface and the faces of the Brillouin zone (Katsnelson \ea 1990),
whereas for Al it is necessary to take into account the contributions of three
body forces. Nevertheless, in describing the phonon spectra of Al
it is impossible to attain an accuracy identical to that for Ir
(Browman \ea 1971, Fomichev \ea 1990). In Ir the contribution of three body
interactions proves to be small (Greenberg \ea 1990).
As a whole, it turns out that surprising as it may seem, Ir is probably
exceeded in the accuracy of the psedopotential description of the lattice
properties only by alkali metals.
This gives hope that the theory can be used to predict those lattice
properties of Ir for which the experimental data are scanty or absent, such as
thermal expansion or the equation of state at different temperatures. To
be sure, the pseudopotential model is oversimplified. Therefore, where it is
possible, the model predictions should be tested against either the experiment
or the results of the {\it ab initio \/} band calculations. Despite the
fact that no simple model can completely substitute the {\it ab initio \/}
calculations, to obtain a sufficiently reliable physical information
about the complex properties of metals even with the help of a model approach
appears to be of interest. At the same time, fully {\it ab initio \/}
calculations of the phonon contributions to the thermodynamical properties
of metals seem for the present to be impossible (or, in any case, are rather
cumbersome). Therefore, the use of the pseudopotential model for these
purposes, where it is possible seems to be justified. This paper presents
the results of the calculations of the coefficient of thermal expansion
at finite temperatures, and the dependence of the elastic moduli on pressure
in terms of the pseudopotential model developed previously for Ir and Rh
(see I). To understand to what extent the simple model used is reliable at
high pressures the electron energy spectrum of Ir and its equation of state
at $T=0$ were calculated {\it ab initio \/}. The results of the calculations
show that the equation of state calculated in the framework of the
pseudopotential model seems to be sufficiently reliable.

\newpar {2} {Thermal expansion and Gruneisen parameters of Ir and Rh}

In calculating the thermal expansion and the Gruneisen parmeters $\gamma(T)$
at finite temperatu\-res as well as the dependencies of the elastic moduli on
pressure $B_{ik} (p)$ use was made of the pseudopential model constructed in
I and giving the optimal description of the phonon spectra. The results of
the calculations in terms of an earlier model (Ivanov \ea 1994) are given
for comparison. The local pseudopotentials $V_{ps}(q)$ used in the two
models are closely similar. However a slight change in the form of
$V_{ps}(q)$  was made in I to describe better the phonon spectra. The
screening function was taken in both cases in the Geldart and Taylor (1970)
approximation (see, in more detail, Greenberg \ea 1990). The temperature
dependence of the volume per atom
$\Omega (T) \equiv\Omega_0 +\Delta\Omega (T)$ ($\Omega_0 =\Omega (T=0)$),
the coefficient of thermal expansion
$$ \alpha_p= {1 \over \Omega} \left ( {\partial \Omega \over \partial T}
\right )_p \eqno (1) $$
and the Gruneisen parameter $\gamma(T)$ are determined by the following
expressions (Vaks \ea 1978):
$$ {\Delta \Omega (T)  \over \Omega_0 } =- {p_{ph} (T) \over B_T},
\eqno (2) $$
$$\alpha_p= {1 \over \Omega B_T}\sum_\lambda \left ({\hbar \omega_\lambda
\over T } \right )^2 N_\lambda (1+N_\lambda ) \gamma_\lambda \eqno(3) $$
$$ \gamma(T)= {\Omega B_T \alpha_p \over C_V(T)} =
{\displaystyle  \sum_\lambda \left ({\hbar \omega_\lambda
\over T } \right )^2 N_\lambda (1+N_\lambda ) \gamma_\lambda \over
\displaystyle \sum_\lambda \left ({\hbar \omega_\lambda
\over T } \right )^2 N_\lambda (1+N_\lambda ) } \eqno (4) $$
where
$$ p_{ph} ={1 \over \Omega} \sum_\lambda \hbar\omega_\lambda N_\lambda
\gamma_\lambda \eqno(5) $$
is the phonon pressure; $\lambda \equiv {\bf q} \nu$, {\bf q} is the
phonon wave vector (sweeping the Brillouin zone); $\nu$ is the branch
number; $\omega_\nu$ is the phonon frequency;
$$ \gamma_\lambda =- {\partial \ln \omega_\lambda \over \partial \ln \Omega}
\eqno (6) $$
are the microscopic Gruneisen parameters;
$$ N_\lambda ={1 \over \displaystyle \exp {\hbar \omega_\lambda \over T} -1 },
\eqno (7) $$
$B_T$ is the bulk modulus at a constant temperature; $C_V(T)$ is the
phonon heat capacity at a constant volume. The microscopic Gruneisen
parameters are calculated by the formula
$$ \gamma_\nu ({\bf q} ) =-{ \Omega \over 2 \omega^2_{\nu {\bf q}}}
\sum_{\alpha\beta} (e^\alpha_{\nu {\bf q}})^* {\partial D_{\alpha\beta}
({\bf q}) \over \partial \Omega} e^\beta_{\nu {\bf q}} \eqno (8) $$
where $D$ is the dynamic matrix; ${\bf e}_{\nu {\bf q}}$ are the polarization
vectors; $\alpha$ and $\beta$ are the Cartesian indices. In our calculations
allowance was also made for the quasiharmonic corrections:
$$
\alpha =\alpha_1 \left [ 1 -\left ( {\Delta \Omega \over \Omega} \right )_1
{\partial \ln B \over \partial \ln \Omega} \right ]
$$
$$
{\Delta \Omega \over \Omega} = \left ( {\Delta \Omega \over \Omega} \right )_1
\left [1+  \left ( {\Delta \Omega \over \Omega} \right )_1
\left (1- {\partial \ln B \over \partial \ln \Omega} \right ) \right ]
\eqno (9)
$$
where index ``1'' means the corresponding expression taken in the lowest order
of anharmonicities. The procedure of calculating the thermal expansion of
metals and its related quantities was discussed in more detail in Vaks \ea
1978.

The results of the calculations are presented in figs. 1--4 and table 1.
First of all, it should be noted once more that $\gamma_\nu ({\bf q})$
vary essentially over the Brillouin zone. This indicates again that the
Gruneisen approximation $\gamma_\nu ({\bf q}) = const$ is completely
unsuited for actual systems. As the temperature increases the macroscopic
Gruneisen parameter $\gamma(T)$ reaches quickly (at $T\ge 0.1 \Theta$,
where $\Theta$ is the Debye temperature) a constant high temperature value.
Alkali metals exibit the analogous behavior (Vaks \ea 1978). However, unlike
alkali metals having $\gamma \approx 1$, in Ir and Rh, as in majority of
transition metals, $\gamma \approx 2$. Note also that the calculated values
of $\gamma (T) $ are rather highly sensitive to the form of the
pseudopotential used (compare the different curves in figs 3b and 4b).

\newpar {3} {Equation of states. Used formulas and approximations}

The equation of state for a metal, that is, the dependence of the pressure
$p$ on $\Omega$ and $T$ is determined by the expression
$$ p(\Omega,T) =p_0(\Omega)+ p_{zp}(\Omega) +p_{ph} (\Omega,T)+
\Delta p_e (\Omega,T) \eqno(10) $$
where
$$ p_0(\Omega)= -{\partial E(\Omega) \over \partial \Omega }, \eqno(11) $$
$E(\Omega)$ is the total energy of the crystal at $T=0$;
$$ p_{zp} =- {\partial \over \partial \Omega} \sum_\lambda \left ( {\hbar
\omega_\lambda \over 2} \right ) ={1 \over 2 \Omega} \sum_\lambda \hbar
\omega_\lambda \gamma_\lambda \eqno (12) $$
is the pressure relating to the zero-point energy; $p_{ph}$ is the phonon
pressure (5);
$$ \Delta p_e ={ \pi^2 \over 6} T^2 {\partial N(E_F) \over \partial \Omega}
\eqno (13) $$
is the temperature dependent contribution to the electron pressure;
$N(E_F)$ is the density of states at the Fermi level $E_F$. We shall neglect
the anharmonic contributions having the same order of magnitude as the
deviation of the $\Delta p_e(T)$ from the ``low-temperature'' ($T\ll E_F$)
expression (13). Generally speaking, the allowance for the contribution of
the zero-point vibrations $p_{zp} (\Omega)$ to the  equation of state in the
scheme with a pseudopotential whose parameters are fitted, among other
things to the condition $\Omega=\Omega^{exp}$ at $p=0$ requires to refit
the parameters. However the results of Vaks \ea 1977 show that even in metals
as light as Li and Na such refitting is not very essential. In addition,
in transition metals, in particular in Ir, the choice of the    value for
the effective charge is not quite  unumbiguous, and strictly speaking,
$Z$ can be dependent of $\Omega$ ($Z\equiv 1$ in alkali metals). Because
we ignore this knowingly more important effect, the results obtained in
refitting the pseudopotential parameters with allowance for $p_{zp}$ are not
presented. The direct calculation shows that that the change in the
results with such a refitting is actually small.

In calculating the major contribution to the pressure $p_0 (\Omega)$ by
(11) in the framework of the pseudopotential model for the energy
$E(\Omega)$, use was made of the expression derived in the second order of
the perturbation theory from a pseudopotential with the exchange-correlation
energy according to analitical approximation (Perdew and Zunger 1981) of the
results of the calculation by the quantum Montne-Carlo method (Ceperly and
Alder 1980). For more details see Greenberg \ea 1990. The {\it ab initio
\/} calculations of the band structure and the total energy at different
$\Omega$ were done by the augmentned plane wave method. The detailed
procedure of these calculations is described by Sigalas \ea 1992 and
Sigalas and Papaconstantopulos 1994.

\newpar {4} {Equation of state. Results of calculations}

The strict calculation of equation of state for a metal by Eq. (10) is
rather difficult. Whereas the electron contributions $p_0$ and $\Delta p_e$
can be calculated form ``the first principles'', the calculation
of the phonon contributions $p_{zp}$ and $p_{ph}$ in the fully ``first-
principle'' approach is a very cumbersome procedure, and no corresponding
results are known to us at that time. For this reason, to calculate the
phonon contributions to pressure requires to use some model concepts.

We'll discuss first the results of the $T=0$ calculations. Fig. 5 shows the
values of $p_0(\Omega)$ calculated in the framework of the pseudopotential
model and the band approach that we used. It is seen that the difference
of the two curves is not too great, at least, at moderate pressures. The
reason is that the values of the bulk modulus: 3.76 Mbar calculated with
the used variant of the band theory (Sigalas \ea 1992) and 2.93 Mbar in the
pseudopotential model (see I) do not differ too much.

Table 2 lists the contribution of the zero point vibrations to the equation
of state at $T=0$. It is seen that this contribution is negligible. The
results of the calculation of the phonon pressure in the pseudopotential
model are shown in figs. 6 and 7. The temperature dependent electron
contribution to the pressure was derived from the results of the band
calculations according to (13). It appears to be very small in comparison
with the phonon contribution (50 times smaller at the temperature
295 K).
The resulting equation of state for Ir at different
temperatures is presented in fig. 8. The results for Rh are similar
and not presented here. Unfortunately,
at present we  have no experimental data to be compared with the above
results.

\newpar {5} {Elastic moduli at high pressures}

Information about the behavior of the Ir and Rh lattice properties at
different pressures can be obtained in the framework of the simple
pseudopotential model used here. The extraction of such information from
experiments seems today to be difficult. Therefore, we present the
results of the corresponding calculations for Ir (fig. 9). The results
for Rh are similar.

\newpar {6} {Conclusion}

The reasons why the lattice properties of Ir and Rh are so succesfully
described in the framework of the used simple pseudopotential model are
not fully understood. However it is believed that if not sufficient, then
at least necessary condition for this is the absence of any noticable
singularities in the density of electron states in the immediate vicinity
of $E_F$. The results of the calculations made by us at different pressures
(figs. 10,11) show that at not too
high pressures this remains the feature of Ir,
and, what's more, the character of the electron structure near $E_F$ is
unchanged. Therefore, it can be assumed that whatever the reason for the
succesful pseudopotential description at $p=0$, the situation at high
compressions will not be too different from that at $p=0$. As a result, one
can expect that the information about the lattice properies of Ir (and its
analogue Rh) at $p\ne 0$ which was derived from the pseudopotential
calculations appears ot be raliable.

The investigation described in the present paper became possible in part
thanks to the financial support of the International Science Foundation
(grant RGQ300) and the Russian Foundation for Fundamental Research
(grant 95-02-06426).

\newpage

\def\r#1#2#3#4#5{\hangindent 10 mm %
\noindent {\sc #1}, 19{#2}, {\it #3}, {\bf #4},
{#5}.}
\def\rrr#1#2#3#4#5{\hangindent 10 mm %
\noindent {\sc #1}, 19{#2}, {\it #3} {\bf #4}
{#5}.}

\centerline{\sc References} \vskip 5mm

\r{Browman, E. G., and Kagan, Yu. M.} {74} {Uspekhi Fiz. Nauk} {112} {369}

\r{Browman, E. G., Kagan, Yu. M., and Kholas, A.} {71}
{Zh. Eksp. Teor. Fiz.} {61} {737}

\r{Ceperly, D. M., and Alder, B. J.} {80} {Phys. Rev. Lett.} {45} {369}

\r{Fomichev, S. V., Katsnelson, M. I., Koreshkov, V. G., and Trefilov, A. V.}
{90} {Phys. Stat. Sol. (b)} {161} {153}

\r{Geldart, D. J. W., and Taylor, R.} {70} {Canad. J. Phys.} {48} {155}

\r{Gornostyrev, Yu. N., Katsnelson, M. I., Mikhin, A. G., Osetskii, Yu. N.,
and Trefilov, A. V.} {94} {Fizika metall. Metallovede.} {77(2)} {79}

\r{Greenberg, B. A., Katsnelson, M. I., Koreshkov, V. G., Osetskii, Yu. N.,
Peschans\-kikh, G. V., Trefilov, A. V., Shamanaev, Yu. F., and Yakovenkova
L. I.
} {90} {Phys. Stat. Sol. (b)} {158} {441}

\r {Ivanov, A. S., Katsnelson, M. I., Mikhin, A. G., Osetskii, Yu. N.,
Rumyantsev, A. Yu., Trefilov, A. V., Shamanaev, and Yu. F., Yakovenkova, L. I.}
{94} {Phil. Mag.} {B69} {1183}

\rrr {Katsnelson, M. I., Naumov, I. I., Trefilov, A. V., Khlopkin, M. N., and
Khromov, K. Yu.} {96} {submitted to Phil. Mag. B} {}{}

\r{Katsnelson, M. I., Peschanskikh, G. V., and Trefilov, A. V.} {90}
{Fiz. Tverd. Tela} {32} {570}

\rrr{Korenovsky, N. L. and Polyakova, V. P.} {90} {in Single crystals of metals
{\rm Moscow, Nauka, in Russian,}} {} {p72}

\r{Perdew, J. R., and Zunger, A.} {81} {Phys. Rev.} {B23} {5048}

\r{Sigalas, M., and, Papaconstantopulos, D. A.} {94} {Phys. Rev.} {B50}
{7255}

\r{Sigalas, M., Papaconstantopulos, D. A., and Bacalis, N. C.} {92}
{Phys. Rev.} {B45} {5777}

\r{Vaks, V. G., Kravchuk, S. P., and Trefilov, A. V.} {77}
{Fiz. Tverd. Tela} {19}
{1271}

\r{Vaks, V. G., Zarochentsev, E. V., Kravchuk, S. P., Safronov, V. P.,
and Trefilov, A. V.} {78} {Phys. Stat. Sol. (b)} {85} {749}

\newpage

\begin{center}
{\sc List of tables to the paper by Katsnelson
\ea ``Thermal expansion \dots''} \end{center} \vskip 5mm

Table 1. The values of the phonon frequencies $\omega({\bf q})$ and the
microscopic Gruneisen parameters $\gamma({\bf q})$  in the simmetric
points points of the Brillouin zone for Ir and Rh. $\omega({\bf q})$ in THz.
\hbox{A-H} stands for the values calculated with the Animalu-Heine pseudopotential, I stands for the
values calculated with the addition proposed in I.

Table 2. The contribution of the zero point vibration to energy and pressure
for Ir and Rh.
A-H stands for the values calculated with the Animalu-Heine pseudopotential, I stands for the
values calculated with the addition proposed in I.
\newpage

\begin{center}
{\sc Figure captions to the paper by Katsnelson
\ea ``Thermal expansion \dots''} \end{center} \vskip 5mm

Fig. 1 Microscopic Gruneisen parameters for Ir calculated for the pseudopotential from
Ivanonv \ea 1994 (dashed line) and from I (solid line).

Fig. 2 Microscopic Gruneisen parameters for Rh calculated for the pseudopotential from
Ivanonv \ea 1994 (dashed line) and from I (solid line).

Fig. 3 Temperature dependence of the linear coefficient of thermal
expansion (a) and the macroscopic Gruneisen parameter (b) for Ir.
Dashed line - calculation for the pseudopotential from Ivanov \ea 1994;
solid line --- calculation for the pseudopotential from I. $\diamond$ in
fig. 3a --- experiment from {\sc Korenovsky and Polyakova, 1990}.
  $T_{pl}=\hbar\omega_{pl}=
\hbar(4 \pi Z e^2 /M \Omega_0) ^{1/2}$, $\omega_{pl}$ is the ionic plasma
frequency.

Fig. 4 Temperature dependence of the linear coefficient of thermal
expansion (a) and the macroscopic Gruneisen parameter (b) for Rh.
Dashed line --- calculation for the pseudopotential from Ivanov \ea 1994;
solid line --- calculation for the pseudopotential from I. $\diamond$ in
fig. 4a --- experiment from {\sc Korenovsky and Polyakova, 1990}.
$T_{pl}=\hbar\omega_{pl}=
\hbar(4 \pi Z e^2 /M \Omega_0) ^{1/2}$, $\omega_{pl}$ is the ionic plasma
frequency.

Fig. 5 Equation of state for Ir at $T=0$. Solid line --- calculation for
the pseudopotential from I; $\diamond$ --- {\it ab initio \/} calculations.

Fig. 6 Dependence of the phonon contribution to pressure upon volume
at $T=295 K$ and $T=T_m$ for Ir (solid line) and Rh (dashed line).
The calculations were made for the pseudopotential from I.

Fig. 7 Temperature dependence of the phonon contribution to pressure for
Ir (solid line) and Rh (dashed line).
The calculations were made for the pseudopotential from I.
$T_{pl}=\hbar\omega_{pl}=
\hbar(4 \pi Z e^2 /M \Omega_0) ^{1/2}$, $\omega_{pl}$ is the ionic plasma
frequency.

Fig. 8 Equation of state for Ir at T=295 K and $T=T_m$. $\diamond$ is {\it ab initio \/}
calculation of $p_0$ with pseudopotential calculation of $p_{ph}$ using the addition proposed in I
at temperature 295 K. $\circ$ is the same at the temperature $T=T_m$.
Solid line --- all quantities were calculated with the pseudopotential proposed in I, temperature
is 295 K. Dashed line --- the same at the temperature $T=T_m$.

Fig. 9 Dependence of the elastic moduli $B_{ik}$ on volume for Ir.
The calculations were made for the pseudopotential from I.

Fig. 10 The density of electron states (a) and the electron energy
spectrum (b) in Ir at equilibrium volume.

Fig. 11 Density of electron states in Ir at different pressures.
\newpage

table 1 \vskip 5mm

\begin{tabular} {|l||c|c|c||c|c|c|}
\hline
& \multicolumn {3} {c||} {$\omega({\bf q})$} & \multicolumn {3} {c|}
{$\gamma({\bf q})$} \\
\cline{2-7}
      &(1,0,0)&(1,1/2,0)&(1/2,1/2,1/2)&(1,0,0)&(1,1/2,0)&(1/2,1/2,1/2) \\
\hline
Ir A-H  &8.06   &7.00     &8.11         &2.39   &2.27     &2.33          \\
        &5.65   &5.85     &4.03         &2.18   &1.83     &1.60          \\
\hline
Ir I    &6.98   &5.90     &7.44         &2.90   &3.01     &2.55          \\
        &5.16   &5.22     &4.01         &2.89   &2.17     &1.72          \\
\hline
Rh A-H  &10.03  &8.63     &9.97         &2.17   &2.10     &2.15          \\
        &7.04   &7.10     &4.91         &1.99   &1.68     &1.44          \\
\hline
Rh I    &7.62   &6.08     &8.51         &3.12   &3.76     &2.51          \\
        &5.93   &5.64     &4.86         &3.34   &2.37     &1.64          \\
\hline
\end{tabular} \vskip 5mm

table 2 \vskip 5mm

\begin{tabular} {|c|c|c|c|c|}
\hline
                    & Ir A-H   & Ir I    & Rh A-H    & Rh I      \\
\hline
$E_{zp}/E_0$$10^4$  & 2.49     & 2.53     & 4.54     & 3.78       \\
\hline
$P_{zp}/P_0$$10^3$  & 2.63     & 3.1      & 3.77     & 5.25       \\
\hline
\end{tabular}
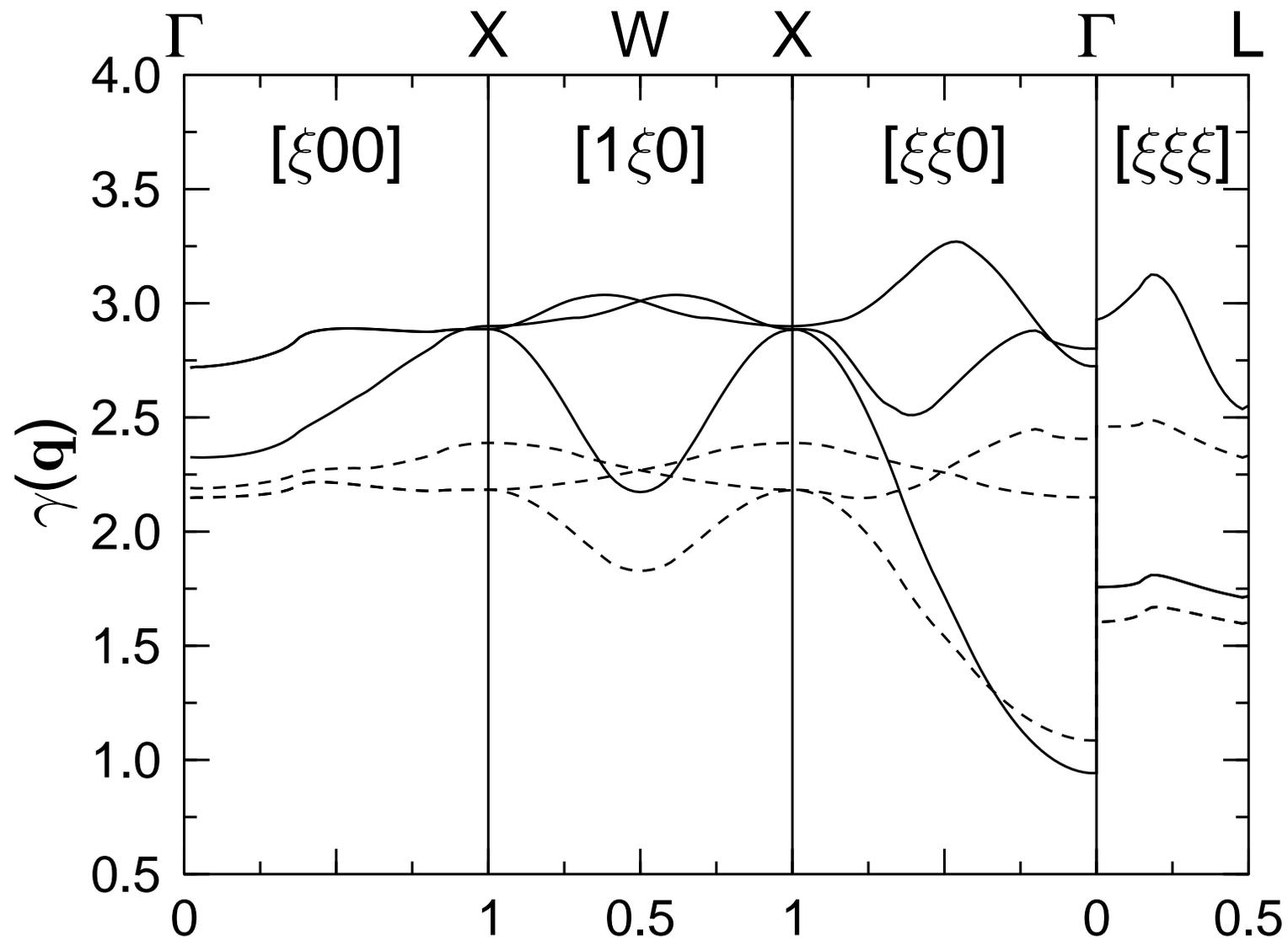
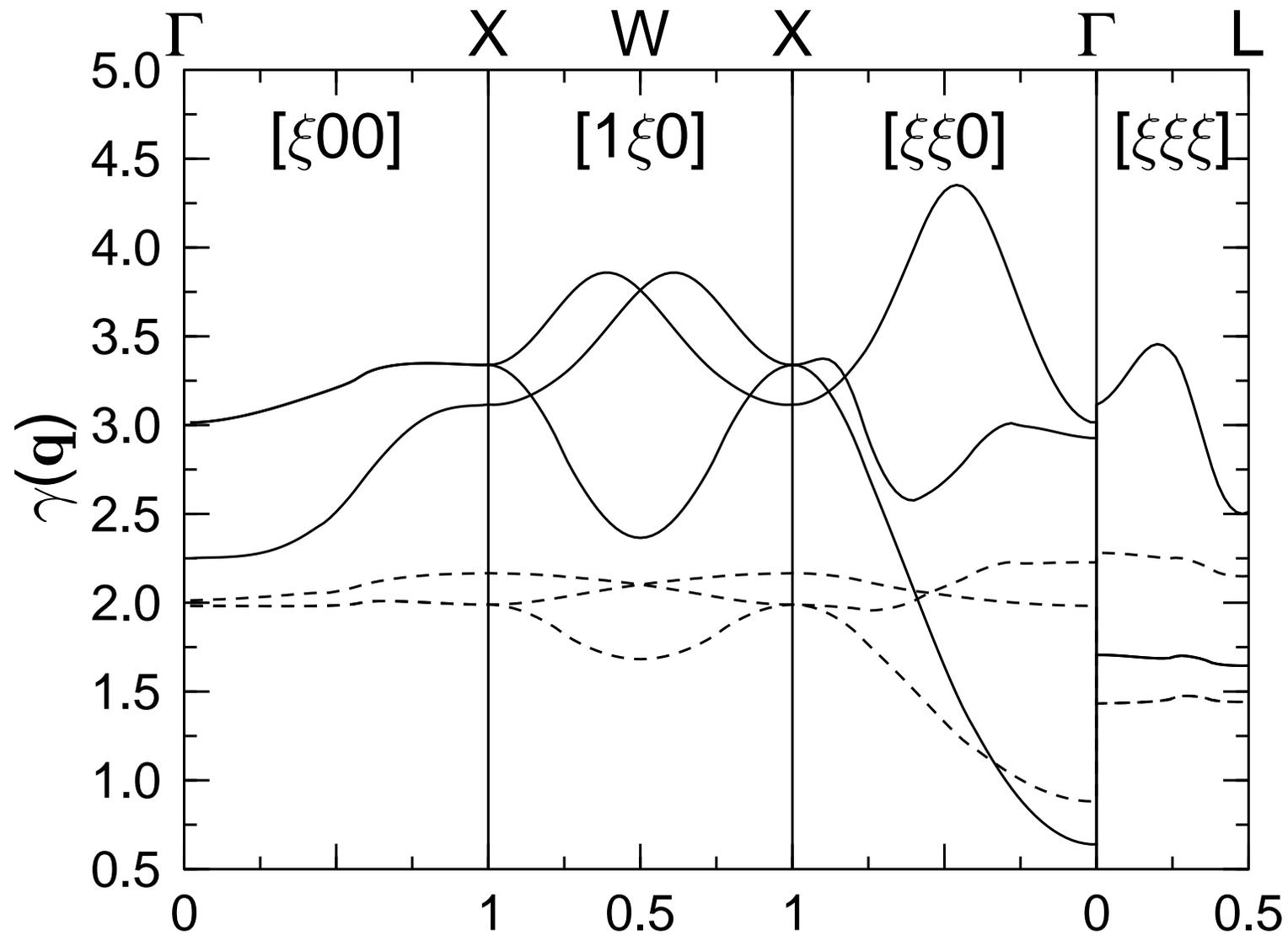
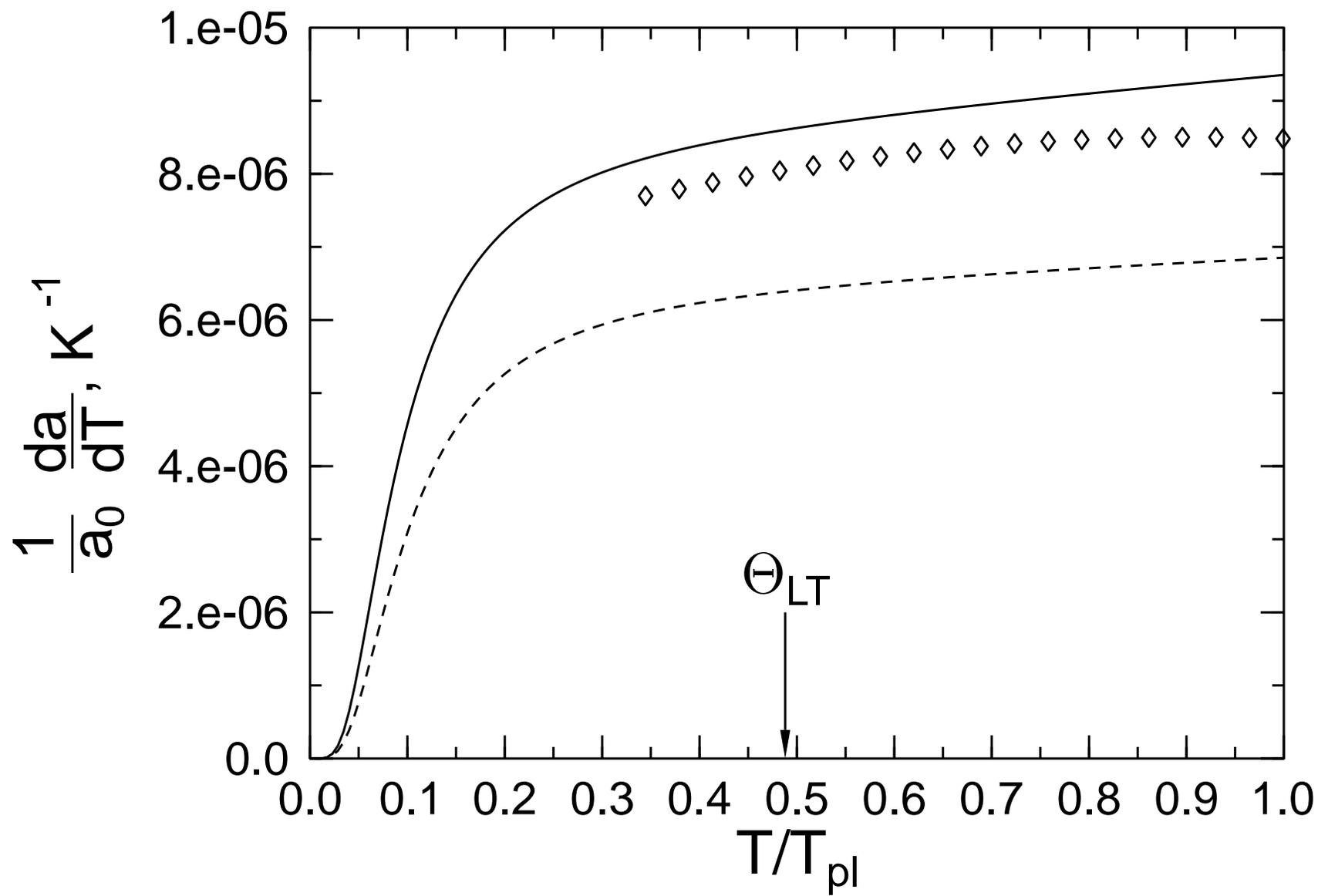
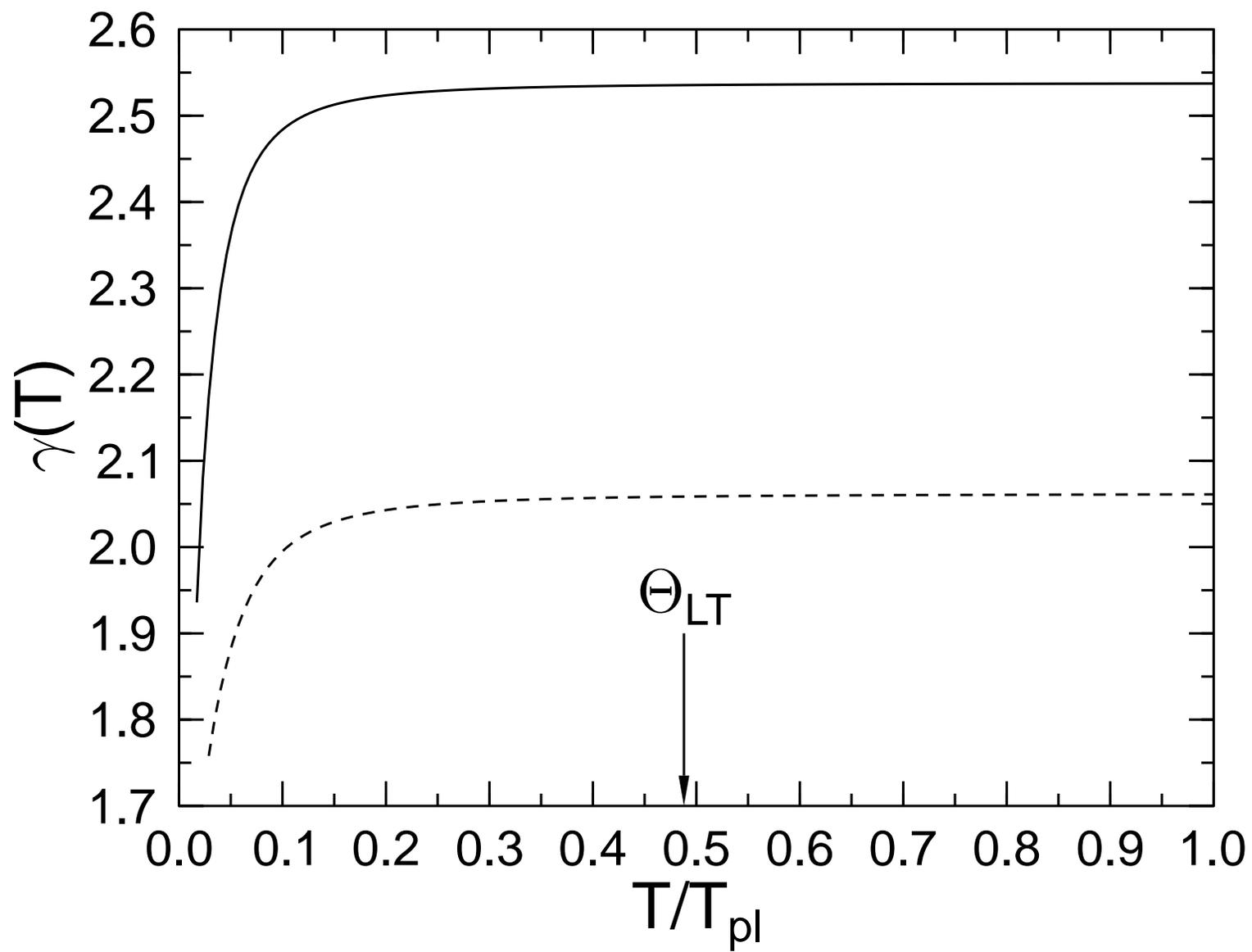
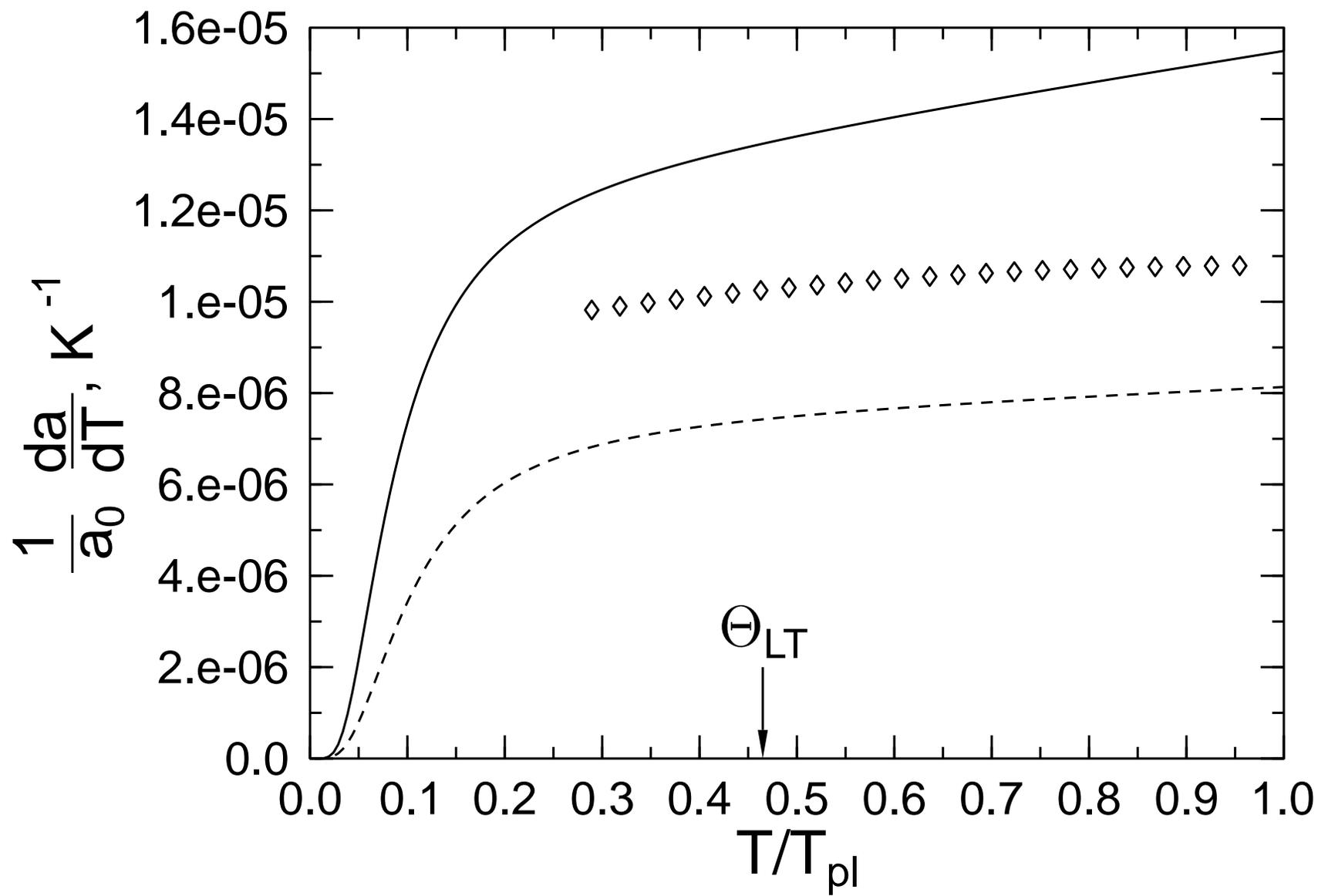
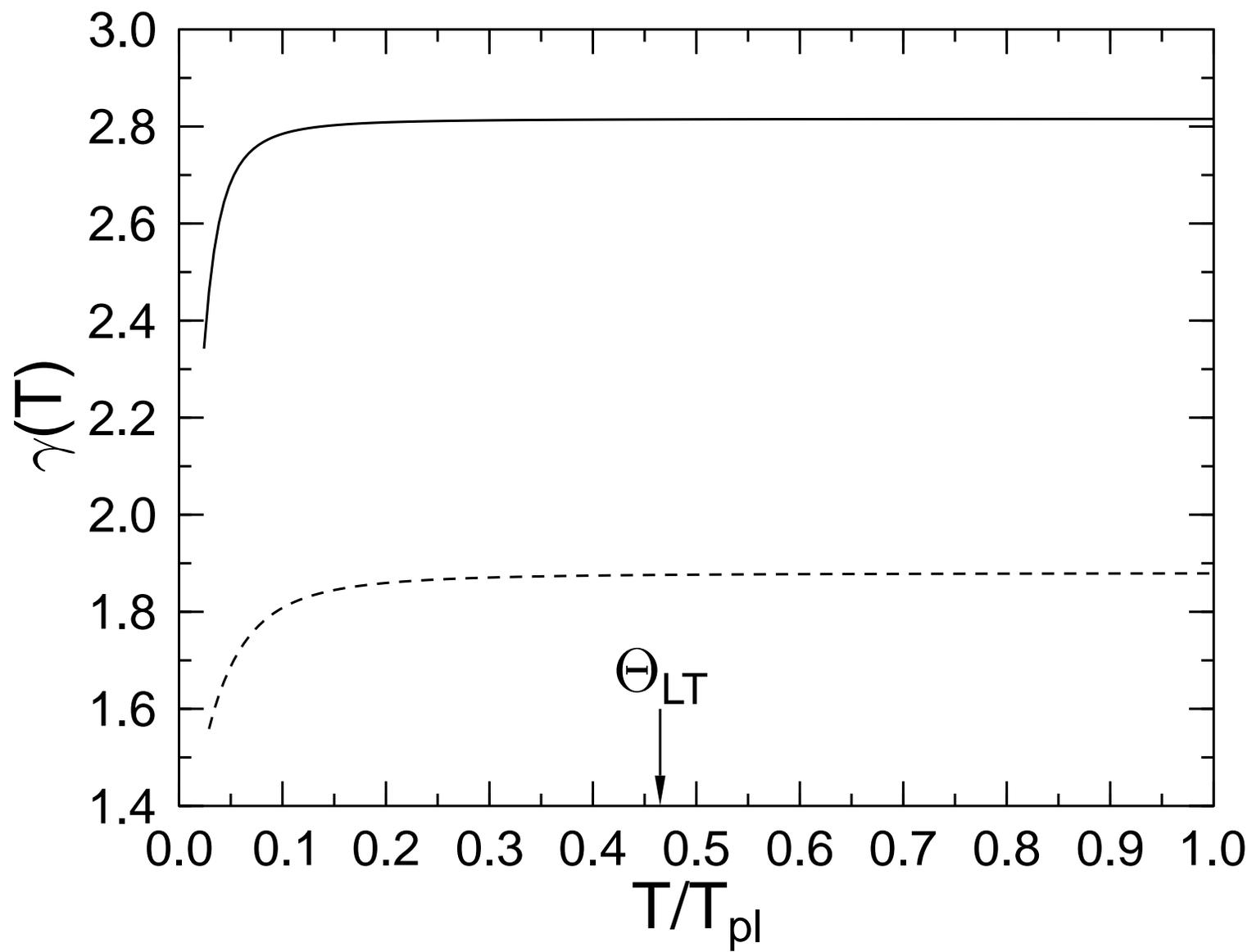
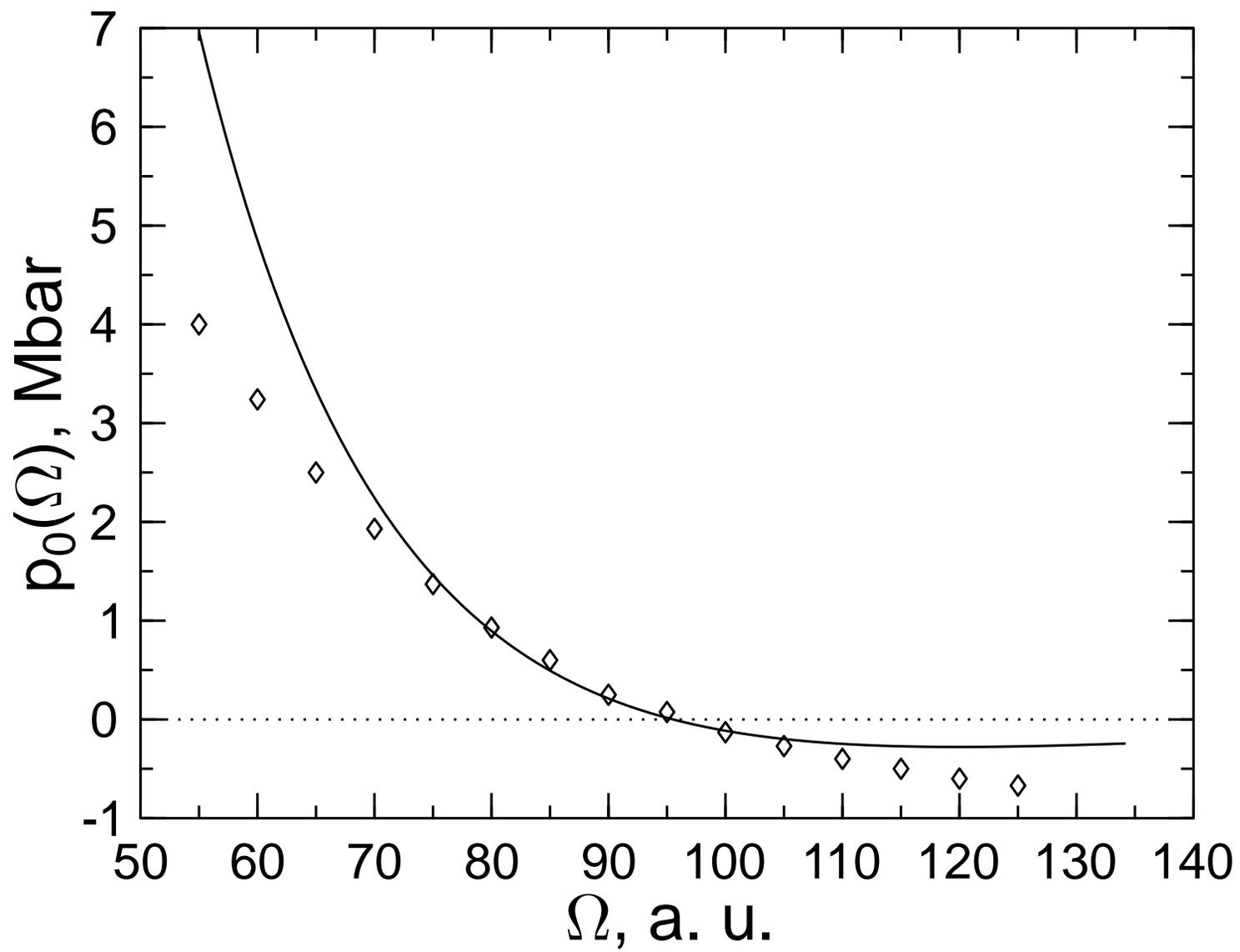
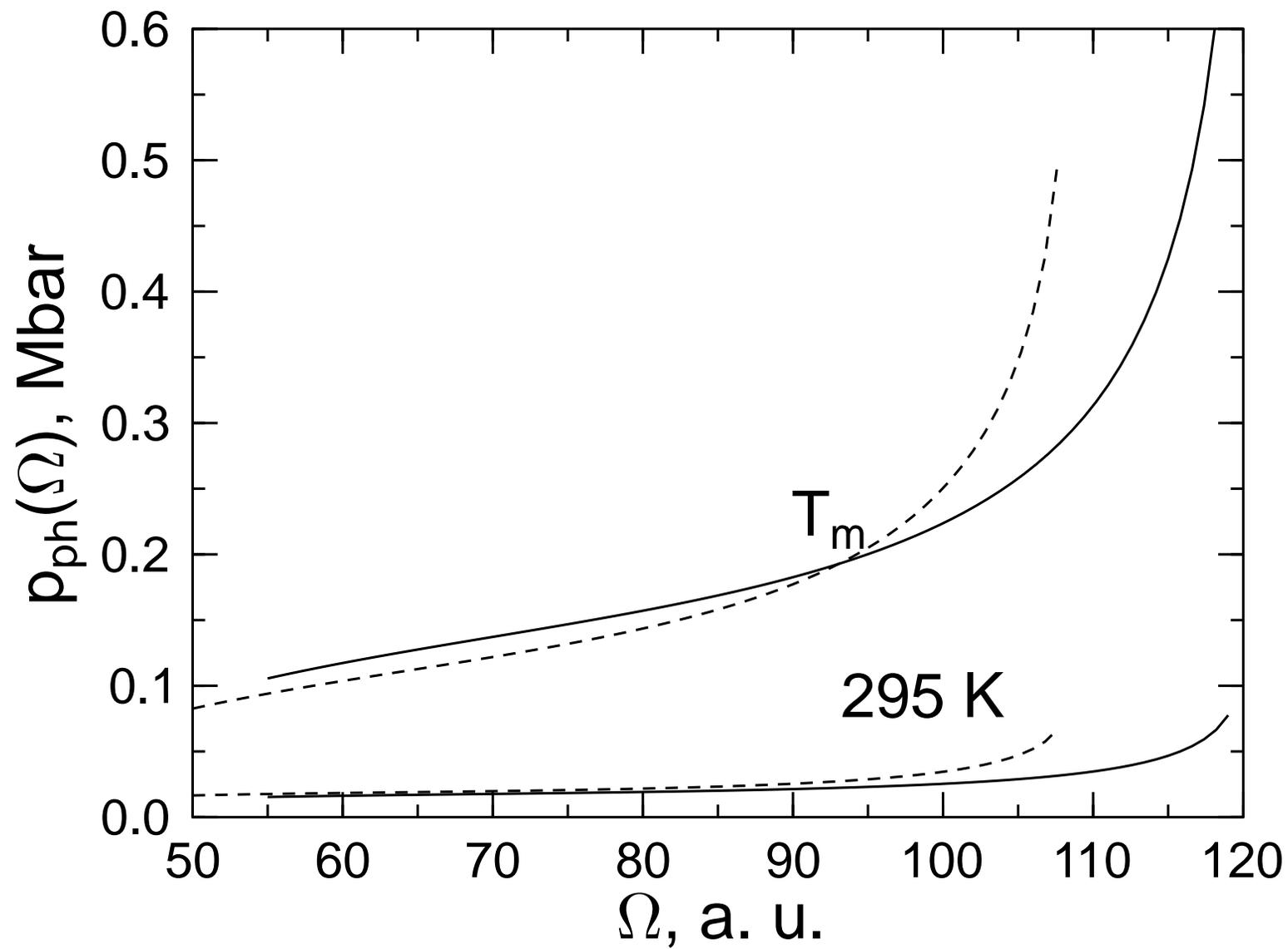
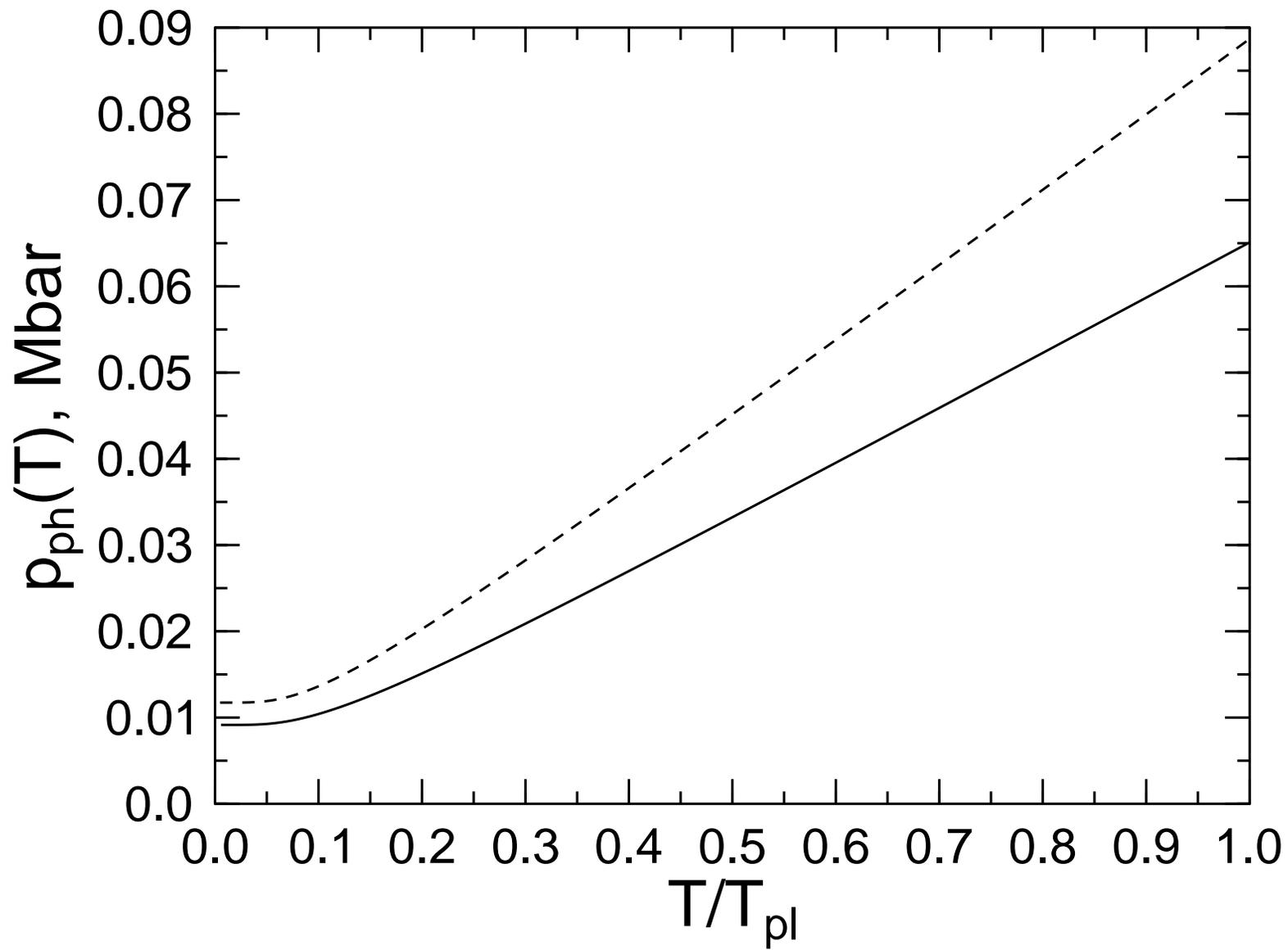
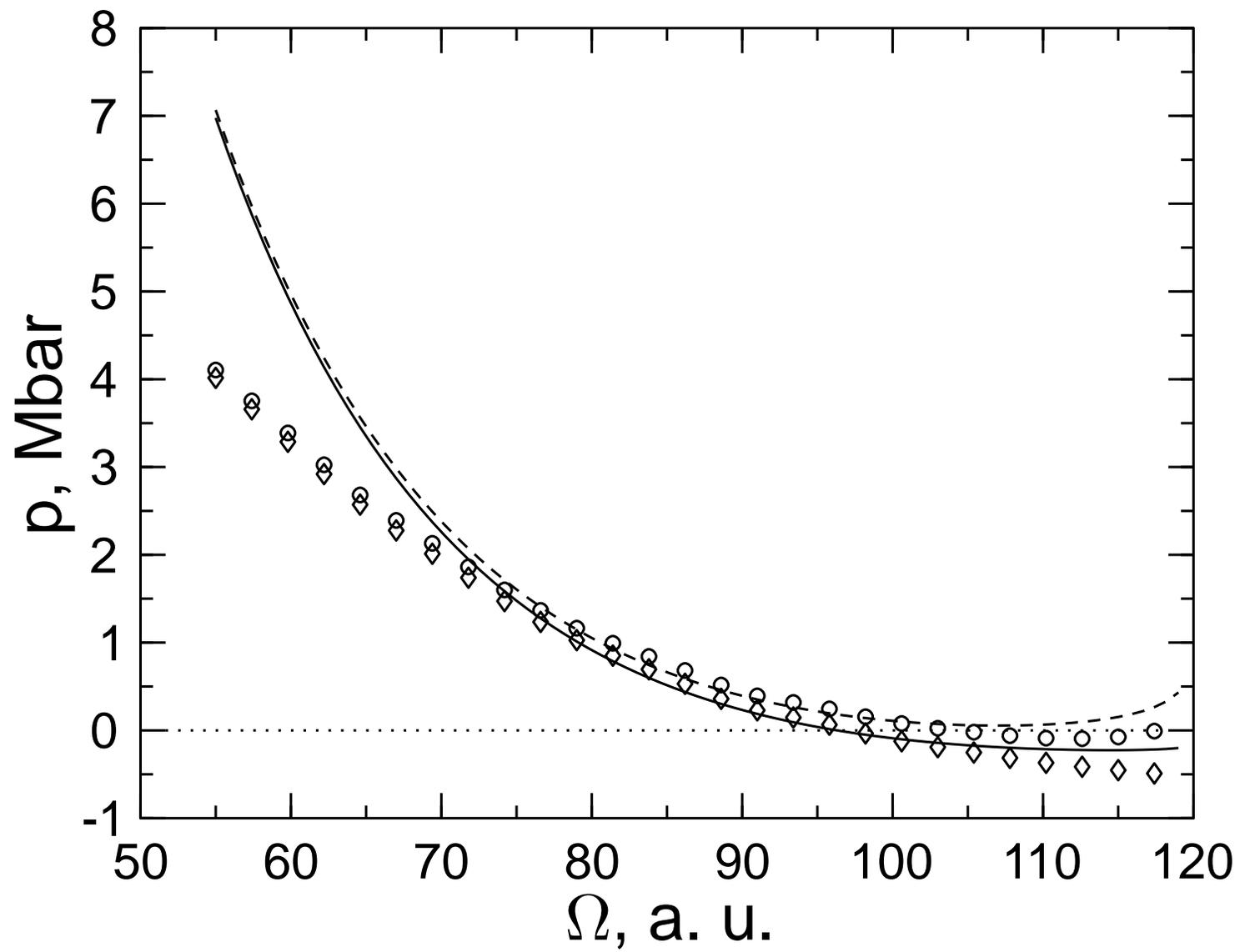
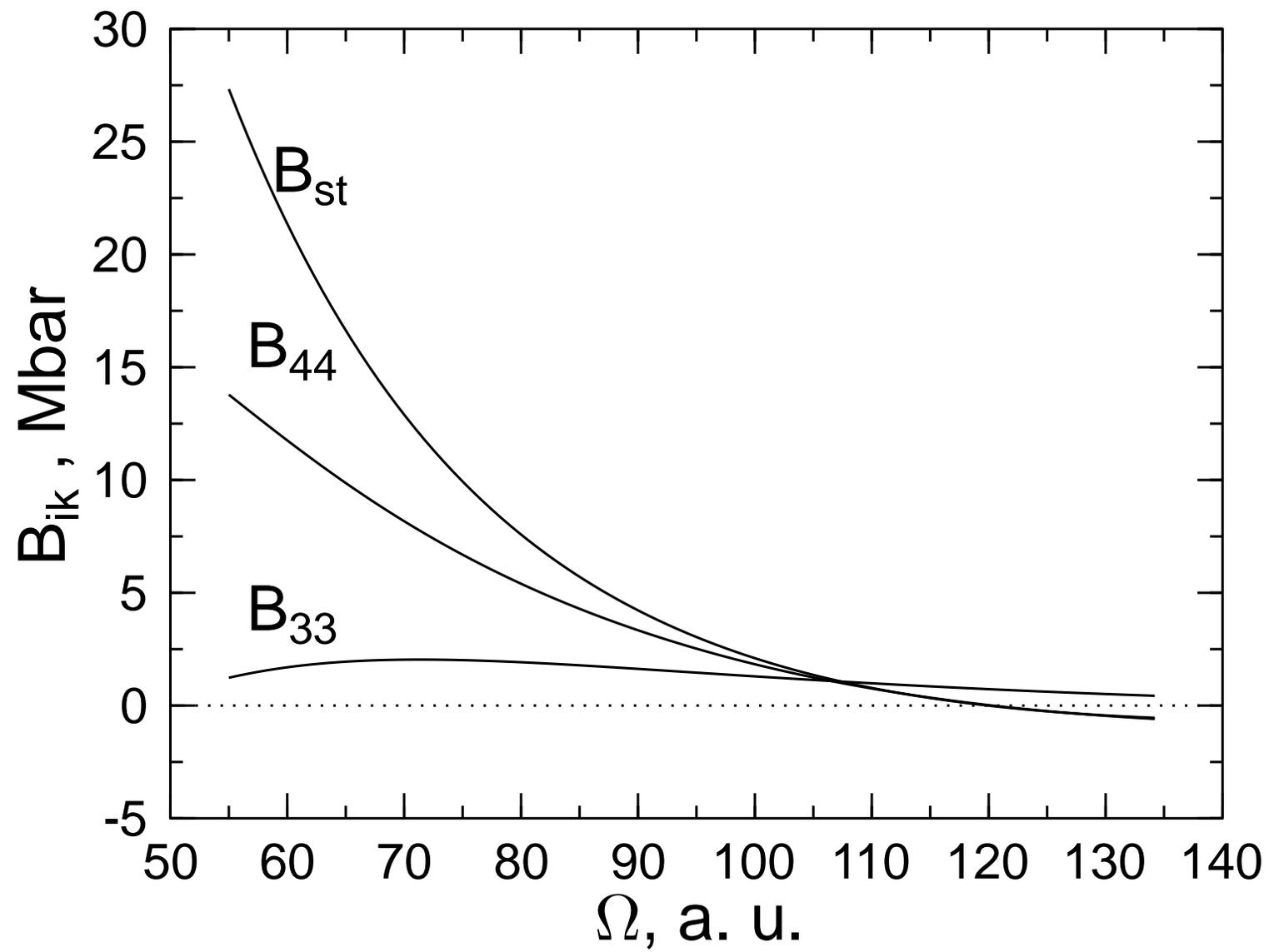
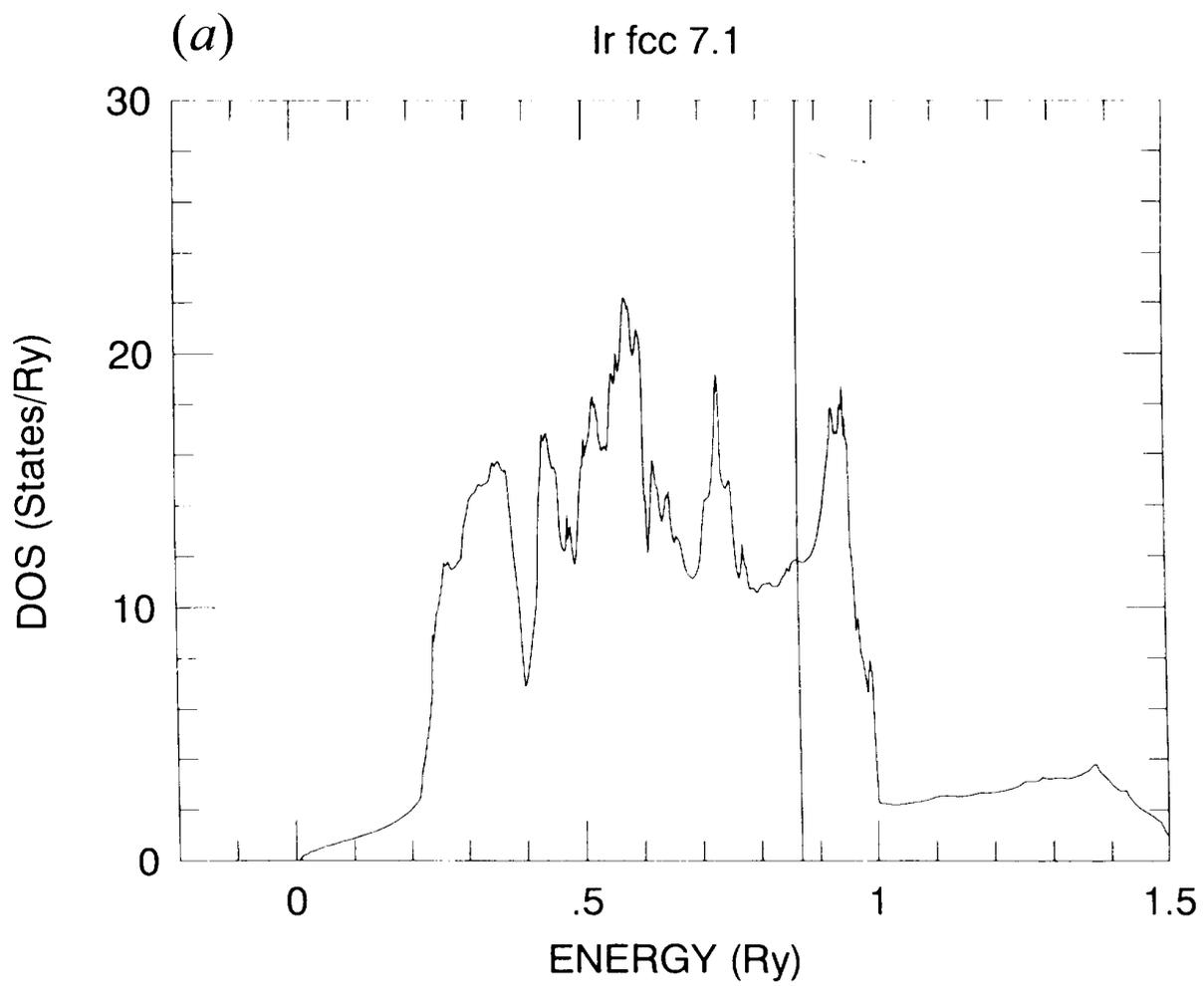
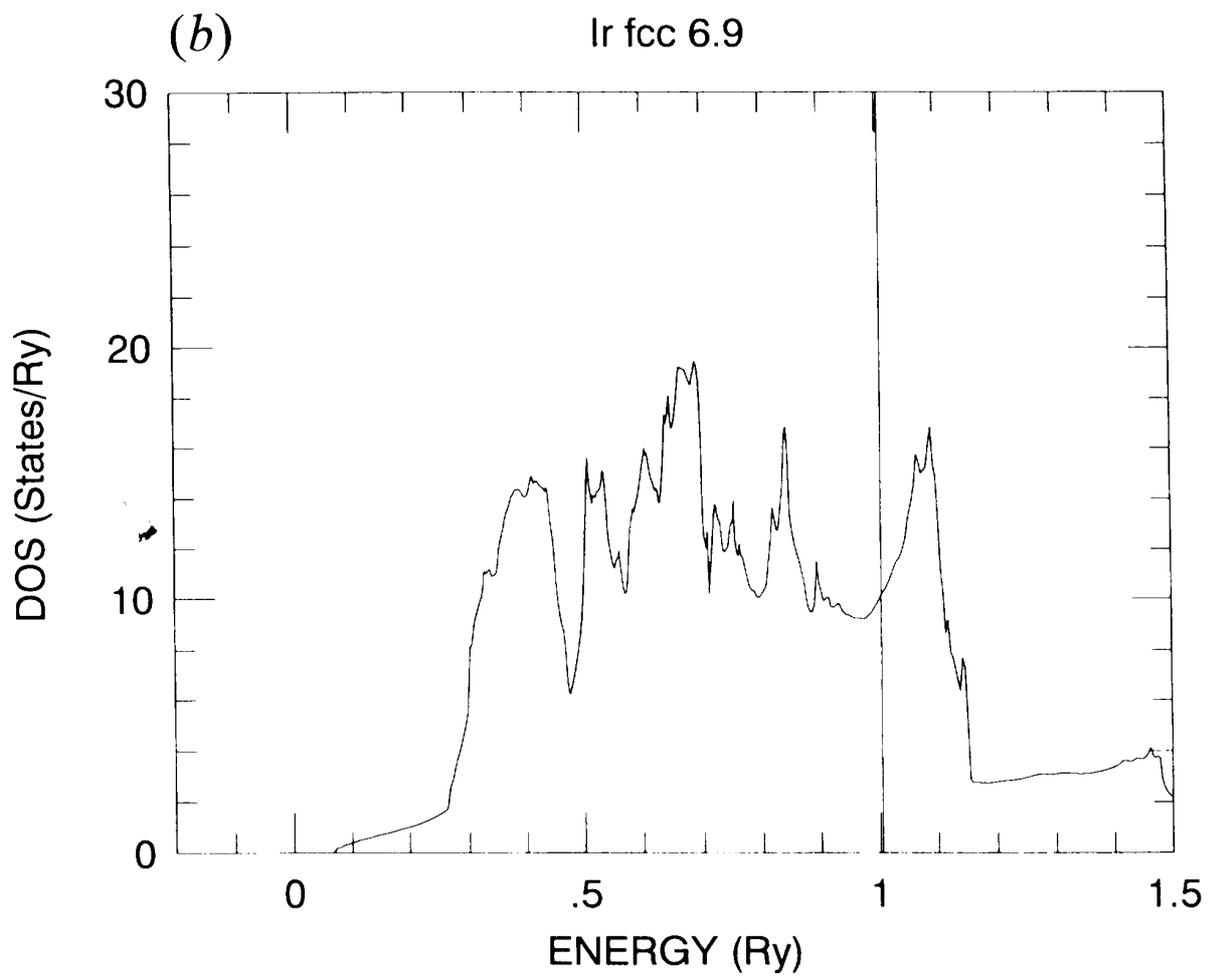
\end{document}